\documentclass[a4paper,twoside]{article}

\usepackage{epsfig}
\usepackage{subcaption}
\usepackage{calc}
\usepackage{amssymb}
\usepackage{amstext}
\usepackage{amsmath}
\usepackage{amsthm}
\usepackage{multicol}
\usepackage{pslatex}
\usepackage{soul}      
\usepackage{xcolor}    
\sethlcolor{blue}    
\usepackage{algorithm2e}
\usepackage[bottom]{footmisc}
\usepackage{natbib}
\usepackage{SCITEPRESS}     

\begin{document}

\title{Enhancing Continual Learning for Software Vulnerability Prediction: Addressing Catastrophic Forgetting via Hybrid‑Confidence‑Aware Selective Replay for Temporal LLM Fine-Tuning}

\author{\authorname{Xuhui Dou\sup{1}, Hayretdin Bahsi\sup{2}\orcidAuthor{0000-0001-8882-4095} and Alejandro Guerra-Manzanares\sup{3}\orcidAuthor{0000-0002-3655-5804}\thanks{ \scriptsize{Corresponding author: alejandro.guerra@nottingham.edu.cn}}}
\affiliation{\sup{1}School of Computer Science, University of Nottingham, Nottingham, United Kingdom}
\affiliation{\sup{2}School of Informatics and Computing, 
Northern Arizona University, Flagstaff, United States of America}
\affiliation{\sup{3}School of Computer Science, University of Nottingham, Ningbo, China}
}


\abstract{Recent work applies Large Language Models (LLMs) to source-code vulnerability detection, but most evaluations still rely on random train--test splits that ignore time and overestimate real-world performance. In practice, detectors are deployed on evolving code bases and must recognise future vulnerabilities under temporal distribution shift. This paper investigates continual fine-tuning of a decoder-style language model (microsoft/phi-2 with LoRA) on a CVE-linked dataset spanning 2018--2024, organised into bi-monthly windows. We evaluate eight continual learning strategies, including window-only and cumulative training, replay-based baselines and regularisation-based variants. We propose Hybrid Class-Aware Selective Replay (Hybrid-CASR), a confidence-aware replay method for binary vulnerability classification that prioritises uncertain samples while maintaining a balanced ratio of VULNERABLE and FIXED functions in the replay buffer. On bi-monthly forward evaluation Hybrid-CASR achieves a Macro-F1 of 0.667, improving on the window-only baseline (0.651) by 0.016 with statistically significant gains ($p = 0.026$) and stronger backward retention (IBR@1 of 0.741). Hybrid-CASR also reduces training time per window by about 17 percent compared to the baseline, whereas cumulative training delivers only a minor F1 increase (0.661) at a 15.9-fold computational cost. Overall, the results show that selective replay with class balancing offers a practical accuracy--efficiency trade-off for LLM-based temporal vulnerability detection under continuous temporal drift.}

\onecolumn \maketitle \normalsize \setcounter{footnote}{0} \vfill

\section{\uppercase{Introduction}}
\label{sec:introduction}

\noindent Software vulnerabilities remain a critical threat to modern computing infrastructure. Public databases such as the Common Vulnerabilities and Exposures (CVE) list report sustained growth in disclosed flaws, with monthly disclosures increasing from around 400 in 2018 to over 2{,}000 in 2024, echoing earlier evidence that vulnerability volumes and exploitation risks continue to rise~\citep{Theisen2020VulnPrediction}. Traditional static analysis tools struggle under this scale: empirical studies report false-positive rates above 80\% in industrial settings, which limits their usefulness for developers and security teams~\citep{Morrison2015HotSOS}.

Machine-learning-based vulnerability prediction attempts to address these limitations by learning patterns from historical data, ranging from metric-based and history-based models to graph neural networks and code-specific deep architectures~\citep{Zhou2019Devign}. More recently, decoder-style large language models (LLMs) for code, fine-tuned on vulnerability datasets, have shown promising results for function-level vulnerability classification~\citep{Konno2025Taltech}. However, most evaluations still rely on random or mixed train--test splits that disregard time, which can introduce severe data leakage and substantially overestimate real-world performance~\citep{Lyu2022Splitting,Fischer2025DupLeak}. In practice, models are deployed on continuously evolving code bases where the distribution of vulnerable and non-vulnerable functions changes over time, a form of concept drift~\citep{Hinder2024ConceptDrift}. Under such drift, static offline training quickly becomes obsolete.

Continual learning (CL) offers a principled way to update models under distribution shift by incrementally training on new data while attempting to preserve performance on previously seen tasks~\citep{Wu2024CLSurvey}. A central challenge in CL is \emph{catastrophic forgetting}, where performance on earlier tasks degrades as the model adapts to new data. For vulnerability detectors deployed over years of evolving code, practical CL strategies must therefore control forgetting while remaining adaptable to new vulnerability patterns. Yet, the interaction between CL strategies and temporal vulnerability detection with LLMs is still poorly understood.

In this paper we formulate temporal vulnerability detection as a binary function-level classification problem, where each function is labelled as either vulnerable or fixed. We construct a CVE-linked dataset from CVEfixes~\citep{Bhandari2021CVEfixes} and related sources, covering disclosures from 2018 to 2024 and grouped into bi-monthly windows. A parameter-efficient decoder LLM, microsoft/phi-2, is adapted using low-rank adaptation (LoRA)~\citep{Hu2021LoRA} and then updated across 42 forward-chained windows. The setting exposes three intertwined challenges observed in practice: catastrophic forgetting when models are updated only on recent windows, strong and time-varying class imbalance between vulnerable and fixed functions, and computational constraints that limit exhaustive retraining on cumulative data.

Building on this formulation, our study addresses the following research questions, adapted from the thesis version of this work:
\begin{itemize}
    \item \textbf{RQ1:} How do different continual learning strategies compare when applied to a parameter-efficient decoder LLM (phi-2 with LoRA) for temporal vulnerability detection?
    \item \textbf{RQ2:} What impact does temporal window granularity (from monthly to annual windows) have on forward prediction performance and stability?
    \item \textbf{RQ3:} To what extent can continual learning strategies mitigate catastrophic forgetting while maintaining plasticity for new vulnerability patterns?
    \item \textbf{RQ4:} What computational trade-offs arise from different continual learning approaches, and how do these affect deployment feasibility in realistic single-GPU environments?
\end{itemize}

To answer these questions, we evaluate eight continual learning strategies, including window-only fine-tuning, cumulative training, replay-based methods and regularisation-based variants, all applied to the same phi-2 backbone. The main contributions of the paper can be summarised as follows:
\begin{enumerate}
    \item \textbf{Temporal protocol.} We design a deployment-faithful temporal evaluation protocol for CVE-linked code with forward-chained training and lagged backward tests (IBR@1/3/5/6). Models are trained only on vulnerabilities known at time $t$ when predicting those disclosed at $t+1$, avoiding temporal leakage that affects random-split evaluations~\citep{Lyu2022Splitting}.
    \item \textbf{Granularity ablation.} We conduct a systematic temporal granularity ablation (1, 2, 3, 6 and 12 months), showing that different granularities obtain comparable Macro-F1 scores (0.651--0.667). Quarterly windows achieve the best average performance, challenging assumptions that there exists a single optimal temporal segmentation.
    \item \textbf{Hybrid-CASR replay.} We propose \emph{Hybrid-CASR}, a confidence-aware, class-balanced replay method that combines uncertainty-based sampling~\citep{Aljundi2019MIR} with an explicit vulnerable/fixed ratio in the replay buffer. On the bi-monthly setting, Hybrid-CASR attains the highest mean Macro-F1 (0.667), with statistically significant gains over the window-only baseline (0.651, $p=0.026$) and stronger backward retention (IBR@1 up to 0.754).
    \item \textbf{Resource--performance analysis.} We provide a resource-performance analysis across methods, reporting F1 per minute, memory usage and relative speed-up. Cumulative training requires about 138.2 minutes per window compared to 8.7 minutes for the window-only baseline, while Hybrid-CASR achieves a favourable trade-off with F1/min $\approx 0.093$ despite higher memory usage.
\end{enumerate}

Overall, the results present Hybrid-CASR as a practical compromise between accuracy, stability and efficiency for decoder-LLM-based vulnerability detectors deployed under continuous temporal drift. To the best of our knowledge, this is the first work to systematically evaluate LLM-based vulnerability detection under a long-horizon, non-stationary temporal protocol with dozens of real-world windows, establishing a reproducible framework that future work can build upon.

\section{\uppercase{Methodology}}
\label{sec:method}

This section provides a thorough description of the experimental setup used in our work, including the experimental protocol, dataset construction, and the evaluation metrics and stastistical techniques used to analyse the results. 

\subsection{Experimental Setup}
\label{sec:setup}

\subsubsection{Model Architecture and Configuration}
\label{sec:model-config}

\noindent
The experiments use \texttt{microsoft/phi-2}, a 2.7B-parameter decoder-only language model, as the base architecture for vulnerability detection. This model size represents a deliberate trade-off between computational constraints and representational capacity. Smaller models below 1B parameters lack sufficient capacity to capture complex code semantics, while models exceeding 7B parameters require multi-GPU setups incompatible with iterative window-wise training. The 2.7B scale permits comprehensive temporal experiments on a single GPU while maintaining competitive performance on code understanding tasks.

Architecturally, encoder models such as CodeBERT achieve strong performance on static benchmarks through bidirectional attention~\citep{Feng2020CodeBERT}, but recent evaluations found that pretrained encoders degrade under temporal evaluation protocols~\citep{Lu2021CodeXGLUE}. These models process bidirectional context, potentially incorporating future signals that violate forward-looking protocols. Decoder architectures process sequences causally, aligning more naturally with the temporal prediction setting where future code is genuinely unseen. Recent work also shows that decoder-based LLMs maintain more consistent performance across code granularities under temporal evaluation~\citep{Konno2025Taltech}, further supporting the choice of a decoder architecture.

Parameter-efficient fine-tuning via Low-Rank Adaptation (LoRA) enables model specialisation while preserving computational efficiency~\citep{Hu2021LoRA}. LoRA is applied to both attention and MLP projection matrices with rank $r=16$, sufficient to capture task-specific patterns without overfitting individual windows. The scaling factor $\alpha=32$ and dropout rate $p=0.05$ are selected based on preliminary experiments on a held-out validation period, balancing adaptation capacity against regularisation. Backbone parameters remain frozen throughout all runs, ensuring consistent representations across temporal windows while allowing task-specific adaptation through low-rank updates. This design aligns with recent parameter-efficient continual learning methods that mitigate catastrophic forgetting~\citep{Qiao2024BotherLess}.

\begin{table}[t]
  \centering
  \caption{Training configuration maintained across all experimental conditions.}
  \label{tab:config}
  \resizebox{\columnwidth}{!}{%
    \begin{tabular}{ll ll}
      \hline
      \textbf{Component} & \textbf{Setting} & \textbf{Component} & \textbf{Setting} \\
      \hline
      Base model & \texttt{microsoft/phi-2} & Learning rate & $2\times10^{-4}$ \\
      Classification head & Binary (2 logits) & Weight decay & 0 \\
      LoRA rank $r$ & 16 & Training epochs & 3 \\
      LoRA $\alpha$ & 32 & Batch size (default) & 32 \\
      LoRA dropout & 0.05 & Batch size (granularity) & 4 \\
      Optimizer & AdamW & Evaluation frequency & Per epoch \\
      Decision threshold & 0.5 (fixed) & Numerical precision & FP32 \\
      \hline
    \end{tabular}%
  }
\end{table}

All experiments are executed on Google Colab Pro with a single NVIDIA A100 GPU (40\,GB memory). The software stack comprises PyTorch~2.0, HuggingFace \texttt{transformers}~4.35, and \texttt{peft}~0.6 for LoRA. Package versions are pinned to ensure consistent behaviour across runs. Resource monitoring tracks wall-clock time and peak memory consumption per training window, enabling cost--benefit analysis when comparing computationally intensive cumulative training against more efficient continual learning approaches.

\subsection{Dataset Construction and Temporal Design}
\label{sec:dataset}

\subsubsection{Data Sources and Quality Control}
\label{sec:sources}

\noindent
The dataset derives from Common Vulnerabilities and Exposures (CVE) records spanning 2018--2024, linked to corresponding fixing commits through the CVEfixes database~\citep{Bhandari2021CVEfixes}. CVEfixes automatically identifies commits that reference CVE identifiers in commit messages or issue trackers, establishing traceability between vulnerability disclosures and code changes. This temporal scope captures recent evolution in vulnerability patterns while providing sufficient historical data for temporal analysis. \citet{Ponta2019Dataset} showed that manually curated CVE--commit links achieve higher precision but limited coverage, motivating automated linking for large-scale temporal studies. CVEfixes contains vulnerabilities from multiple programming languages. In this study we focus on function-level instances from the dominant languages in the corpus (primarily C/C++), filtering out samples from other languages to keep the setting homogeneous.

Each CVE record is validated to ensure data quality. Entries marked as ``Rejected'' in the CVE database are excluded to remove false positives and duplicates. When multiple commits reference a single CVE, priority is given to commits that directly remove vulnerable patterns rather than preparatory refactoring or documentation-only changes.

Temporal anchoring is critical for evaluation validity. Each instance is timestamped using the CVE disclosure date rather than the commit date, reflecting information available to practitioners at decision time. Commit timestamps often precede disclosure by weeks or months during coordinated release processes, creating an information asymmetry that would leak future knowledge into training data. The disclosure-clock protocol therefore ensures that models train only on vulnerabilities known at time $t$ when predicting those disclosed at time $t{+}1$, consistent with temporal evaluation recommendations for code intelligence benchmarks~\citep{Lu2021CodeXGLUE}.

\subsubsection{Function-level Instance Generation}
\label{sec:instances}

\noindent
The preprocessing pipeline extracts function-level instances from repository snapshots. For each CVE--commit pair, two versions are retrieved: the vulnerable function $f^{\text{pre}}$ from the parent revision immediately before the fix, and the patched function $f^{\text{post}}$ from the fixing commit. This before--after pairing yields balanced positive and negative examples:
\begin{itemize}
    \item Vulnerable instances: $(f^{\text{pre}}, y=1)$ representing code that still contains the security flaw;
    \item Fixed instances: $(f^{\text{post}}, y=0)$ representing remediated code after the patch.
\end{itemize}
Multi-file commits generate multiple function pairs when patches span several files or functions. Each function is processed independently, preserving function-level granularity aligned with typical code review units. Functions exceeding the model context window are truncated from the end, potentially losing some context but maintaining consistent input lengths.

Timeline-wide deduplication addresses a critical threat identified by~\citet{Allamanis2019Duplication}, who showed that code duplication can inflate performance by up to 40\% when duplicates cross train--test splits. \citet{Fischer2025DupLeak} further demonstrated that inter-dataset duplication significantly affects LLM evaluation validity. In the CVEfixes setting, near-duplicate functions naturally arise when the same fix is back-ported across maintenance branches or mirrored in repository forks. The deduplication process therefore computes normalised hashes of function bodies (removing whitespace and comments) and retains only the earliest occurrence of each unique function. This guarantees
$\mathcal{D}_{t+1} \cap \bigl(\bigcup_{i \leq t} \mathcal{D}_i\bigr) = \emptyset$,
preventing any function from appearing both in a training window and in a future test window.

\subsubsection{Temporal Window Granularity}
\label{sec:windows}

\noindent
The 2018--2024 timeline is segmented into bi-monthly windows, yielding 42 consecutive periods. This granularity balances several design constraints. Finer granularity (monthly) offers higher temporal resolution but increases variance due to sparse positive examples in low-activity periods. Coarser granularity (quarterly or longer) improves statistical stability but risks mixing heterogeneous vulnerability types across release cycles and security campaigns. The temporal distribution aligns with observations by \citet{Theisen2020VulnPrediction}, who reported increasing disclosure rates in recent years and warned that models trained on older data may not generalise.

\begin{figure*}[t]
\centering
\includegraphics[width=\textwidth]{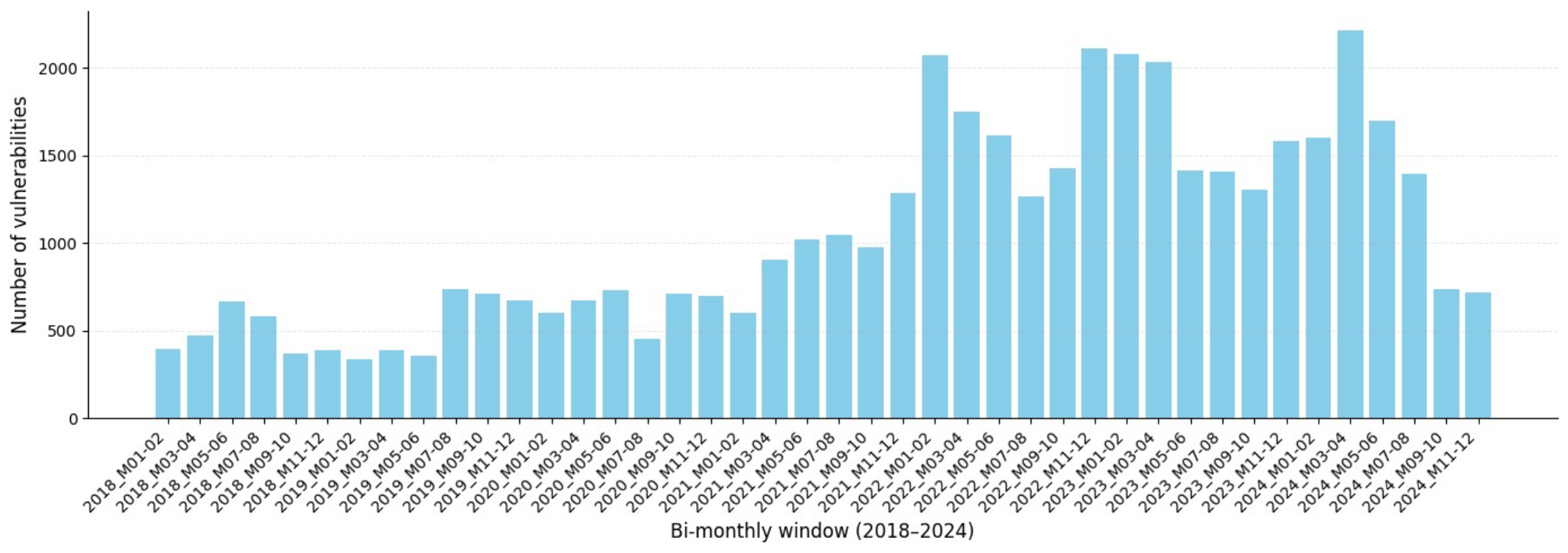}
\caption{Distribution of vulnerability disclosures across bi-monthly windows (2018--2024), illustrating volume growth and temporal variability.}
\label{fig:disclosure-volume}
\end{figure*}

Figure~\ref{fig:disclosure-volume} shows the temporal distribution of vulnerabilities, revealing substantial variation across periods. The data exhibits a marked increase after 2021, with some windows containing over 2{,}000 disclosures compared to fewer than 400 in earlier periods. This growth likely reflects expanded security auditing, automated discovery tools and increased open-source adoption. Beyond volume, shifts in languages, frameworks and dependency practices contribute to concept drift as characterised by \citet{Hinder2024ConceptDrift}, who distinguish between sudden drift (e.g.\ the 2021--2022 transition) and gradual drift within otherwise stable periods.

Bi-monthly windows maintain a median positive-class prevalence of 42\% (IQR 35--49\%), providing sufficient vulnerable examples for stable gradient estimates without extreme imbalance. The median window contains 743 instances (IQR 531--1394), enabling meaningful batch sampling while avoiding the computational burden of very large windows.

\subsection{Experimental Protocol}
\label{sec:protocol}

\subsubsection{Forward Temporal Evaluation Framework}
\label{sec:forward-eval}

\noindent
The evaluation protocol implements strict forward-chaining to simulate realistic deployment. Models trained on window $W_t$ are evaluated exclusively on the subsequent window $W_{t+1}$, mimicking the challenge of predicting future vulnerabilities using only historical knowledge. This one-step-ahead forecasting prevents information leakage and provides an honest assessment of temporal generalisation. The temporal aggregation approach follows the recommendations by~\citet{Lyu2022Splitting}, who showed that data splitting decisions strongly influence reported performance, with temporal splits providing more realistic estimates than random partitioning.

Three baseline configurations establish reference points:

\begin{itemize}
    \item \textbf{Zero-shot baseline.} The pre-trained phi-2 model is applied without fine-tuning to test whether general code understanding transfers to vulnerability detection. This baseline typically achieves 50--55\% Macro-F1, indicating limited but non-trivial zero-shot capability.
    \item \textbf{Window-only training.} For each window $W_t$ we start from the same pre-trained phi-2 checkpoint and fine-tune it only on the data in $W_t$, discarding both the training data and parameter updates from previous windows. This configuration exposes catastrophic forgetting as performance on earlier windows deteriorates after training on later ones.
    \item \textbf{Cumulative training.} The model is fine-tuned on all historical data up to window $t$,
    $\mathcal{D}^{\text{cum}}_t = \bigcup_{i=1}^{t} W_i$.
    This computationally expensive baseline approximates an upper bound on performance.
\end{itemize}

\subsubsection{Backward Retention Assessment}
\label{sec:backward-eval}

\noindent
Beyond forward prediction, the protocol evaluates backward retention to quantify knowledge preservation. After training on window $W_t$, models are tested on previous windows $W_{t-k}$ for $k \in \{1,3,5,6\}$. These backward evaluations reveal how effectively continual learning strategies maintain performance on historical vulnerabilities while adapting to new patterns.

The Immediate Backward Retention (IBR) metric at lag $k$ measures F1 on $W_{t-k}$ after training through $W_t$. Comparing IBR across lags reveals the ``half-life'' of learned knowledge under different strategies. Methods that achieve high forward accuracy and strong backward retention offer a better stability--plasticity trade-off, addressing the dilemma highlighted in continual learning surveys~\citep{VandeVen2022CLSurvey}.

\subsubsection{Continual Learning Strategies}
\label{sec:cl-strategies}

\noindent
Six continual learning approaches are evaluated on top of the LoRA-based backbone, in addition to the baselines above.

\textbf{Replay-based methods} maintain buffers of historical examples to rehearse during training on new windows. These methods build on iCaRL~\citep{Rebuffi2017iCaRL}, which established rehearsal as an effective forgetting mitigation strategy. Replay-1P stores instances from the immediately preceding window, while Replay-3P maintains three windows of history. When training on $W_t$, current data are interleaved with buffered examples. Buffer management follows a FIFO policy with uniform sampling from retained instances. \citet{Han2023CLCompare} compared multiple rehearsal variants and found buffer size and selection strategy to be key design dimensions.

\textbf{Confidence-Aware Selective Replay (CASR)} prioritises rehearsal of challenging examples identified through model uncertainty. Following the maximally interfered retrieval principle~\citep{Aljundi2019MIR}, examples with maximum class probability below a threshold $\tau=0.7$ or misclassified instances receive priority for retention. This selective approach maintains smaller buffers while focusing on decision boundaries where forgetting most harms performance.

\textbf{Hybrid-CASR} is a novel contribution that combines confidence-based selection with explicit class balancing. In our dataset the FIXED class is substantially more frequent than the VULNERABLE class, so pure uncertainty-based replay, as in CASR, tends to select many more FIXED samples because most uncertain examples come from the majority class. Hybrid-CASR therefore first constructs a class-balanced candidate set by sampling equal numbers of VULNERABLE and FIXED instances from the buffer, and then ranks examples within each class by prediction confidence. The replay buffer is partitioned such that 70\% of slots are filled by high-uncertainty examples selected by these CASR criteria, and the remaining 30\% are drawn uniformly to maintain coverage. This hybrid design ensures that replayed samples are both informative (uncertain) and approximately class-balanced, directly targeting the class-imbalance issue that arises from varying vulnerability rates across temporal windows.

\textbf{LB-CL (Label-Balanced Continual LoRA)} modifies the training objective to account for class imbalance without maintaining replay buffers. Class-weighted cross-entropy with weights inversely proportional to class frequency within each window is applied. This requires no additional memory but can struggle under extreme distribution shift. \citet{Wu2024CLSurvey} emphasise class balancing as a key factor when task distributions vary.

\textbf{OLoRA (Orthogonality-regularised LoRA)} constrains parameter updates to preserve previous knowledge geometrically. Inspired by gradient episodic memory (GEM)~\citep{LopezPaz2017GEM} and A-GEM~\citep{Chaudhry2019AGEM}, which prevent interference by projecting gradients away from previous task directions, OLoRA encourages new LoRA matrices to remain approximately orthogonal to subspaces spanned by past updates. This is implemented via a regularisation term $\beta \mathcal{L}_\perp(A_t, \mathcal{S}_{t-1})$, where $\mathcal{S}_{t-1}$ summarises historical update directions and $\beta=0.1$ balances adaptation flexibility against interference. \citet{Song2023ConPET} work on parameter-efficient continual tuning shows that orthogonality constraints can effectively reduce forgetting in LLMs.

\begin{table}[t]
  \centering
  \caption{Continual learning strategies with mechanisms and memory requirements.}
  \label{tab:cl_strategies}
  \resizebox{\columnwidth}{!}{%
    \begin{tabular}{p{0.13\textwidth} p{0.42\textwidth} p{0.33\textwidth}}
      \hline
      \textbf{Strategy} & \textbf{Mechanism} & \textbf{Memory requirements} \\
      \hline
      Replay-1P   & Uniform sampling from previous window (Rebuffi et al., 2017). &
                    Memory buffer storing samples from the immediately previous window. \\
      Replay-3P   & Uniform sampling from 3 previous windows (Rebuffi et al., 2017). &
                    Memory buffer storing samples from the last 3 windows. \\
      CASR        & Uncertainty-based selective replay (Aljundi et al., 2019). &
                    Memory buffer with prioritised vulnerable samples. \\
      Hybrid-CASR & CASR + class balancing (this work). &
                    Memory buffer with class-balanced sampling across VULNERABLE / FIXED. \\
      LB-CL       & Class-weighted loss (Wu et al., 2024). &
                    No extra memory beyond model parameters. \\
      OLoRA       & Orthogonality constraints on LoRA updates (Lopez-Paz and Ranzato, 2017;
                    Chaudhry et al., 2019). &
                    LoRA adapter parameters (and small statistics for orthogonal subspace). \\
      \hline
    \end{tabular}%
  }
\end{table}

\subsection{Evaluation Metrics and Statistical Analysis}
\label{sec:metrics}

\subsubsection{Evaluation Metrics}
\label{sec:eval-metrics}

\noindent
\textbf{Primary metric.}
Macro-F1 at a fixed decision threshold of 0.5 is adopted as the primary metric. This choice reflects operational priorities where both classes carry equal importance: missing vulnerabilities (false negatives) leads to security risk, while false alarms (false positives) waste developer effort. Unlike accuracy, which can be misleading under class imbalance, Macro-F1 treats both classes equally by averaging their individual F1 scores~\citep{Saito2015PRBetter}.

For a binary label set $\{0,1\}$ with per-class precision $\mathrm{P}_c$ and recall $\mathrm{R}_c$, the per-class F1 is
$\mathrm{F1}_c = \frac{2\,\mathrm{P}_c\,\mathrm{R}_c}{\mathrm{P}_c + \mathrm{R}_c}$,
and the macro average is
$\mathrm{Macro\text{-}F1} = \tfrac{1}{2}\bigl(\mathrm{F1}_{0} + \mathrm{F1}_{1}\bigr)$.

\textbf{Temporal metrics.}
Given the non-stationary nature of vulnerability patterns, additional metrics capture temporal dynamics:
\begin{itemize}
    \item \textbf{Forward F1}: performance on the immediate next window $W_{t+1}$ after training on $W_t$, measuring generalisation to future vulnerabilities;
    \item \textbf{IBR@k}: performance on $W_{t-k}$ after training through $W_t$ for $k \in \{1,3,5,6\}$, quantifying catastrophic forgetting;
    \item \textbf{Stability index}: coefficient of variation of F1 across windows for each method.
\end{itemize}

\subsubsection{Temporal Aggregation and Statistical Testing}
\label{sec:statistics}

\noindent
Performance aggregation across windows uses the arithmetic mean
$\bar{m} = \frac{1}{K}\sum_{t=1}^{K} m_{t+1}$,
where $K$ denotes the number of evaluation windows. Method comparisons employ paired tests on window-wise performance differences. The Wilcoxon signed-rank test accommodates non-normal distributions common in bounded metrics like F1. Effect sizes (Cliff's delta) complement significance tests by quantifying practical importance beyond statistical detectability.

\subsection{Resource--Performance Analysis}
\label{sec:resource-analysis}

\noindent
Computational cost strongly affects deployment feasibility. Each run logs wall-clock training time and peak GPU memory per window, enabling construction of accuracy--efficiency frontiers. The resource--performance profile for method $M$ at window $t$ is defined as
$\mathbf{r}_{t}^{(M)} = \bigl(\text{time}_t^{(M)},\, \text{memory}_t^{(M)},\, \mathrm{F1}_{t+1}^{(M)}\bigr)$.
This three-dimensional characterisation exposes whether gains in Macro-F1 justify additional computational investment. Cumulative training achieves slightly higher average F1 but requires up to 15.9$\times$ longer training time compared to window-only training in later windows.

\section{\uppercase{Results}}
\label{sec:results}

\noindent
This section presents experimental findings from temporal vulnerability detection
experiments conducted between 2018 and 2024. The analysis addresses the four
research questions defined in Section~\ref{sec:introduction} through evaluation of
temporal granularity effects, continual learning strategies, knowledge retention
and computational trade-offs.

\subsection{Temporal Granularity Analysis}
\label{sec:granularity-results}

\noindent
The impact of temporal window granularity on vulnerability detection performance
is evaluated using the window-only LoRA baseline across five time scales. This
analysis addresses RQ2 by examining how temporal segmentation affects both
prediction accuracy and stability. The granularities span from fine-grained monthly
windows to coarse annual aggregations, providing insight into the bias--variance
trade-off inherent in temporal segmentation.

The experimental configuration maintains identical hyperparameters across
granularities to isolate the effect of window size. Models are trained using
\texttt{microsoft/phi-2} with LoRA (rank $=16$, $\alpha=32$, dropout $=0.05$),
three epochs per window, learning rate $2 \times 10^{-4}$ and batch size 8. The
smaller batch size is necessitated by memory constraints when processing highly
variable window sizes.

\begin{table}[t]
  \caption{Temporal granularity impact on vulnerability detection performance. 
  All metrics are computed from forward evaluation on held-out windows.}
  \label{tab:granularity}
  \centering
  \resizebox{\columnwidth}{!}{%
    \begin{tabular}{lccccccc}
      \hline
      Granularity & Windows & Mean F1 & Std Dev & IQR & Min F1 & Max F1 \\
      \hline
      1-month     & 82 & 0.654 & 0.054 & 0.075 & 0.506 & 0.778 \\
      2-month     & 41 & 0.651 & 0.044 & 0.058 & 0.539 & 0.714 \\
      3-month     & 27 & 0.667 & 0.033 & 0.041 & 0.572 & 0.714 \\
      6-month     & 13 & 0.669 & 0.029 & 0.032 & 0.611 & 0.702 \\
      12-month    &  6 & 0.669 & 0.030 & 0.035 & 0.604 & 0.695 \\
      \hline
    \end{tabular}%
  }
\end{table}

Table~\ref{tab:granularity} summarises performance statistics across all temporal granularities. Different temporal granularities achieve comparable mean F1 scores in the range
0.651--0.669. Monthly windows reach F1~$=0.654$ with higher variability
(standard deviation 0.054), reflecting sensitivity to month-specific fluctuations
in vulnerability types and disclosure patterns. Bi-monthly windows yield similar
mean performance (0.651) with improved stability (standard deviation 0.044).

\begin{figure}[t]
\centering
\includegraphics[width=\columnwidth]{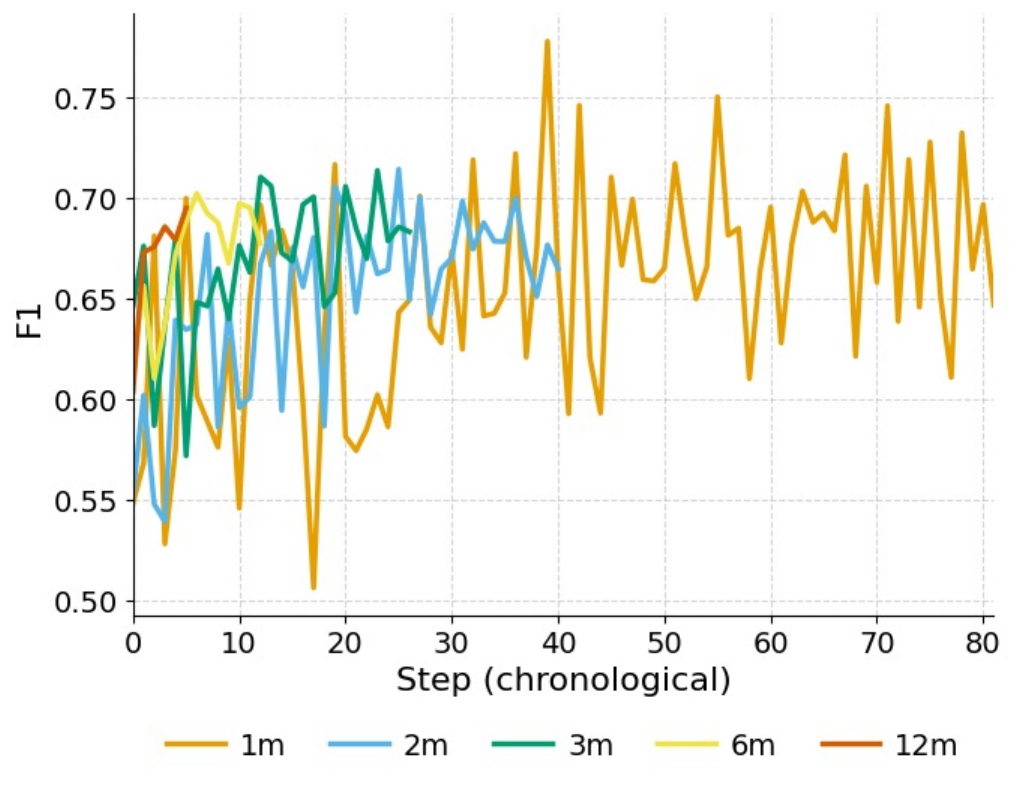}
\caption{Forward one-step F1 scores across different temporal granularities (1, 2, 3, 6, 12 months), showing variability and trends over chronological evaluation windows.}
\label{fig:forward-granularity}
\end{figure}

Figure~\ref{fig:forward-granularity} illustrates forward one-step prediction
performance across granularities. Monthly windows show higher variability with
more extreme peaks and troughs, while longer granularities exhibit more consistent
patterns. Several 1-month windows exceed 0.75 F1, but
drops below 0.55 are also more frequent, confirming the higher variance.

Quarterly, semi-annual and annual windows yield slightly higher mean performance
(0.667--0.669) with progressively lower variance, consistent with the expected
bias--variance trade-off. The small performance variation across temporal granularities (F1 range
0.651--0.669) reveals an unexpected characteristic: models trained at different
time scales achieve remarkably similar effectiveness. This has practical implications: organisations debating update frequencies
can prioritise resource availability rather than
performance optimisation, as quarterly updates (F1~$=0.667$) lose little compared to
monthly retraining (F1~$=0.654$). However, misclassification analysis indicates
that different granularities fail on different vulnerability types, suggesting
that granularity affects \emph{which} vulnerabilities are detected rather than
\emph{how many}.

\subsection{Continual Learning Strategy Comparison}
\label{sec:method-comparison}

\noindent
To address RQ1 and RQ3, eight approaches are evaluated under the standard
bi-monthly protocol: baselines, replay-based strategies, regularisation-based
methods and hybrid techniques. All methods share the same backbone and
hyperparameters (batch size 32 for this comparison), differing only in their
continual learning mechanisms.

The analysis focuses on the intersection of windows where all methods complete
training, ensuring fair comparison. Some configurations fail on specific windows;
those windows are excluded here and left for future investigation.

\begin{table}[t]
\centering
\caption{Forward evaluation performance of continual learning methods (37 bi-monthly windows).}
\label{tab:method-forward}
  \centering
  \resizebox{\columnwidth}{!}{%
    \begin{tabular}{lccccc}
      \hline
      Method & Mean F1 & Std Dev & Min F1 & Max F1 & Win Rate \\
      \hline
      Hybrid-CASR & 0.667 & 0.037 & 0.566 & 0.728 & 46.3\% \\
      Cumulative  & 0.661 & 0.039 & 0.571 & 0.749 & 41.5\% \\
      Replay-1P   & 0.659 & 0.050 & 0.510 & 0.743 & 39.0\% \\
      CASR        & 0.659 & 0.038 & 0.580 & 0.737 & 36.6\% \\
      LB-CL       & 0.651 & 0.044 & 0.534 & 0.728 & 31.7\% \\
      Window-only & 0.651 & 0.044 & 0.539 & 0.714 & 29.3\% \\
      Replay-3P   & 0.622 & 0.045 & 0.497 & 0.698 & 24.4\% \\
      OLoRA       & 0.599 & 0.046 & 0.464 & 0.698 & 17.1\% \\
      Zero-shot   & 0.504 & 0.153 & 0.000 & 0.729 & 0.0\% \\
     \hline
    \end{tabular}%
  }
\end{table}

Table~\ref{tab:method-forward} reports forward evaluation metrics on the 37 bi-monthly windows where all methods succeed. Here, ``mean F1'' denotes the arithmetic mean of Macro-F1 scores across these evaluation windows for a given method. Distinct performance clusters
emerge that reflect different continual learning philosophies.

\begin{figure}[t]
\centering
\includegraphics[width=\columnwidth]{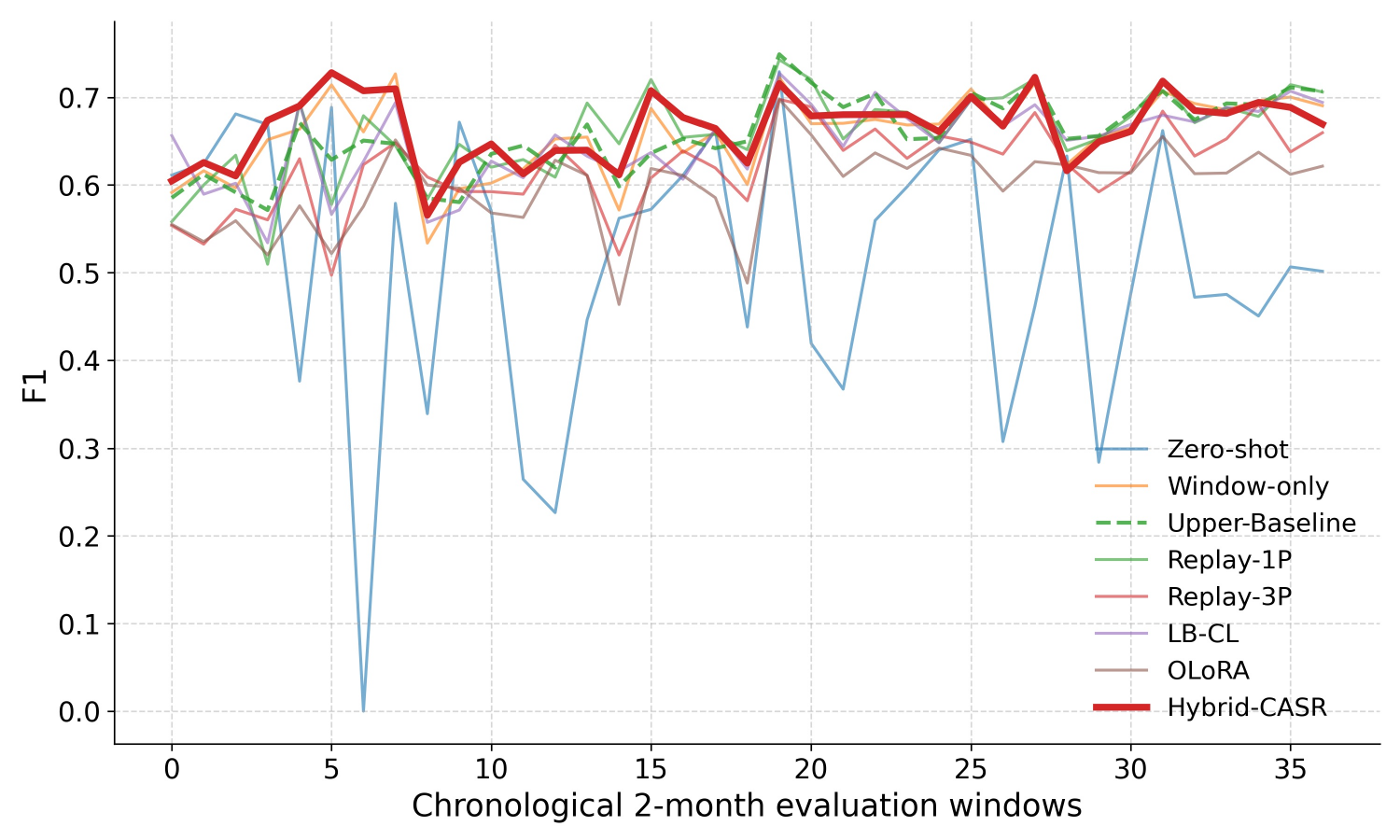}
\caption{Forward F1 comparison across all methods using bi-monthly windows, highlighting differences in performance level and stability over time.}
\label{fig:method-comparison-2m}
\end{figure}

Figure~\ref{fig:method-comparison-2m} visualises forward performance across the
bi-monthly protocol. Hybrid-CASR and other CL methods exhibit much more stable
behaviour than the highly variable zero-shot baseline. The latter displays dramatic
drops to near-zero F1 in certain windows, whereas CL methods maintain performance
floors above 0.4 F1.

Hybrid-CASR emerges as the top performer with mean F1~$=0.667$, achieving
consistent improvements across windows. Its success stems from combining
confidence-based example selection with explicit class balancing, addressing both
catastrophic forgetting and class imbalance. The cumulative baseline attains
comparable mean performance (0.661) but at substantially higher computational
cost (Section~\ref{sec:resource-performance}).

Replay-1P and CASR obtain similar mean F1 (0.659), illustrating the effectiveness
of selective memory retention. The window-only baseline and LB-CL both reach
F1~$=0.651$, serving as reference points without explicit replay.

Methods emphasising strong regularisation show more mixed results. Replay-3P
(0.622) appears to over-constrain the model with excessive historical data, while
OLoRA (0.599) suggests orthogonality constraints can be overly restrictive in this
domain, where vulnerability patterns evolve substantially.

Hybrid-CASR achieves the highest mean F1 (0.667), a 0.016 absolute improvement
(2.5\% relative) over the window-only baseline (0.651). A Wilcoxon signed-rank
test on paired window scores yields $p = 0.026$, indicating statistically significant
gains. The effect size (Cliff's $\delta = 0.103$) corresponds to a small but consistent
advantage.

\subsection{Knowledge Retention Analysis}
\label{sec:retention}

\noindent
Beyond forward prediction, practical deployment requires models to maintain
detection capability for previously encountered vulnerability patterns. The experimental framework therefore
includes backward evaluation at multiple temporal lags, quantifying the
stability--plasticity trade-off inherent in continual learning.

Immediate Backward Retention (IBR) at lag $k$ measures F1 on $W_{t-k}$ after
training up to $W_t$. Four lags are evaluated: IBR@1 (immediate predecessor),
IBR@3, IBR@5 and IBR@6 (one year under bi-monthly segmentation).

\begin{table}[t]
  \centering
  \caption{Knowledge retention measured by Immediate Backward Retention at multiple lags. Decay rate is defined as $(\mathrm{IBR@1} - \mathrm{IBR@6}) / \mathrm{IBR@1}$.}
  \label{tab:retention-detailed}
  \resizebox{\columnwidth}{!}{%
    \begin{tabular}{lrrrrrr}
      \hline
      \textbf{Method} & \textbf{IBR@1} & \textbf{IBR@3} & \textbf{IBR@5} &
      \textbf{IBR@6} & \textbf{Decay rate} & \textbf{AUC}$^a$ \\
      \hline
      Replay-1P   & 0.791 & 0.747 & 0.734 & 0.729 & 7.8\% & 0.750 \\
      Hybrid-CASR & 0.741 & 0.726 & 0.716 & 0.710 & 4.2\% & 0.723 \\
      CASR        & 0.734 & 0.719 & 0.707 & 0.706 & 3.8\% & 0.717 \\
      LB-CL       & 0.718 & 0.703 & 0.691 & 0.687 & 4.3\% & 0.700 \\
      Window-only & 0.713 & 0.701 & 0.689 & 0.693 & 2.8\% & 0.699 \\
      Replay-3P   & 0.702 & 0.688 & 0.676 & 0.673 & 4.1\% & 0.685 \\
      Cumulative  & 0.661 & 0.661 & 0.661 & 0.661 & 0.0\% & 0.661 \\
      OLoRA       & 0.612 & 0.598 & 0.587 & 0.584 & 4.6\% & 0.595 \\
      Zero-shot   & 0.493 & 0.493 & 0.493 & 0.493 & 0.0\% & 0.493 \\
      \hline
    \end{tabular}%
  }
  \vspace{1mm}
  {\footnotesize $^a$Area under the retention curve, normalised to [0,1].}
\end{table}

Table~\ref{tab:retention-detailed} reports retention metrics, revealing distinct
memory profiles across strategies. Replay-1P achieves the highest immediate retention (IBR@1~$=0.791$), with
moderate decay (7.8\%) by lag 6. Hybrid-CASR attains strong retention
(IBR@1~$=0.741$) with lower decay (4.2\%), achieved through selective
preservation of challenging examples and class balancing. The cumulative
baseline maintains perfect stability (0\% decay) but lower absolute retention
(0.661) than most selective methods, suggesting that exhaustive memory may
introduce interference under distribution shift.

\begin{figure}[t]
\centering
\includegraphics[width=\columnwidth]{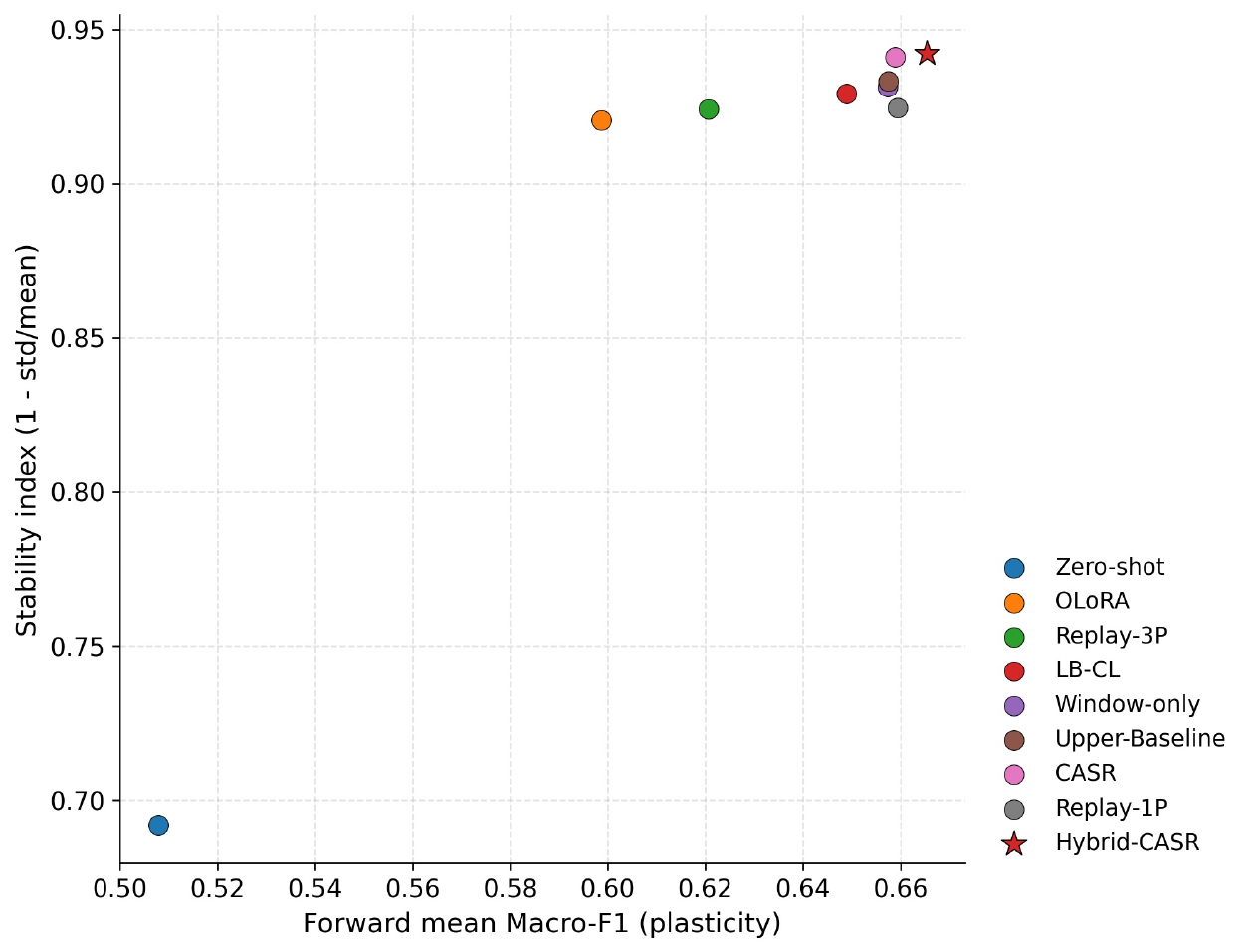}
\caption{Plasticity--stability trade-off, showing the relationship between forward learning capability and performance consistency. Methods in the upper-right region achieve the best balance.}
\label{fig:plasticity-stability}
\end{figure}

The relationship between forward learning (plasticity) and performance
consistency (stability) provides insight into method behaviour under drift.
Figure~\ref{fig:plasticity-stability} maps methods in this space.

Hybrid-CASR occupies the favourable upper-right region, combining high forward
performance with strong stability. This balance arises from its selective design:
confidence-based selection focuses on decision boundaries, while class balancing
prevents collapse toward majority classes.

\subsection{Computational Cost and Efficiency}
\label{sec:resource-performance}

\noindent
Resource consumption heavily influences deployment feasibility, especially in
settings requiring frequent model updates. Table~\ref{tab:resources-detailed}
summarises training time, memory usage and efficiency metrics averaged over 41
common windows.

\begin{table}[t]
  \caption{Resource consumption and efficiency. Metrics are averaged across 
  41 bi-monthly training windows.}
  \label{tab:resources-detailed}
  \centering
  \resizebox{\columnwidth}{!}{%
    \begin{tabular}{lcccccc}
      \hline
      Method & Time (s) & Time (min) & Speedup & GPU (MB) & F1/min & Efficiency \\
      \hline
      Hybrid-CASR &  432 &  7.2 & 1.20$\times$ & 19,847 & 0.093 & 124\% \\
      Window-only &  520 &  8.7 & 1.00$\times$ & 13,525 & 0.075 & 100\% \\
      Cumulative  & 8,291 & 138.2 & 0.06$\times$ & 13,525 & 0.005 &   6\% \\
      \hline
    \end{tabular}%
  }
\vspace{1mm}
{\footnotesize Memory is measured for LoRA fine-tuning in FP32 precision without quantisation. F1/min denotes F1 score divided by training time in minutes.}
\end{table}

Hybrid-CASR achieves higher F1 while requiring less time than window-only
(432\,s vs.\ 520\,s). The F1/min metric quantifies this advantage: Hybrid-CASR
delivers 0.093 F1 points per training minute versus 0.075 for window-only,
a 24\% efficiency gain. This improvement stems from selective replay: focusing
on challenging examples reduces the number of gradient steps required for
convergence. The additional memory (about 47\% more than baseline) reflects
buffer storage and uncertainty tracking.

The cumulative baseline is computationally prohibitive: 15.9$\times$ longer
training (138.2 minutes vs.\ 8.7 minutes) yields only marginal F1 gains. As the
historical dataset grows, this inefficiency worsens, making the approach
impractical for operational settings with daily or weekly updates.

\subsection{Error Analysis and Challenging Scenarios}
\label{sec:error-analysis}

\noindent
Analysis of windows where the window-only baseline performs poorly reveals
systematic patterns in detection difficulty. The bottom quartile of windows by
F1 corresponds to periods of rapid change in disclosure patterns, particularly
around major security events and coordinated campaigns.

Table~\ref{tab:hard-windows} reports results on two of the most challenging
windows, providing insight into robustness under extreme distribution shift. Here, $\Delta$F1 is defined as the difference in Macro-F1 between each method and the window-only baseline on the same window (F1\textsubscript{method} $-$ F1\textsubscript{window-only}), so positive values indicate an improvement over the baseline.

\begin{table}[t]
  \caption{Method performance on challenging temporal windows. 
  Values show F1 and improvement over the window-only baseline.}
  \label{tab:hard-windows}
  \centering
  \resizebox{\columnwidth}{!}{%
    \begin{tabular}{lcccc}
      \hline
      Method & 2019\_M05--06 F1 & $\Delta$F1 & 2020\_M05--06 F1 & $\Delta$F1 \\
      \hline
      Window-only & 0.539 & ---    & 0.548 & ---    \\
      Hybrid-CASR & 0.598 & +0.059 & 0.612 & +0.064 \\
      CASR        & 0.606 & +0.067 & 0.597 & +0.049 \\
      Replay-1P   & 0.584 & +0.045 & 0.588 & +0.040 \\
      Cumulative  & 0.585 & +0.046 & 0.574 & +0.026 \\
      LB-CL       & 0.557 & +0.018 & 0.561 & +0.013 \\
      Replay-3P   & 0.569 & +0.030 & 0.520 & -0.028 \\
      OLoRA       & 0.478 & -0.061 & 0.464 & -0.084 \\
      Zero-shot   & 0.412 & -0.127 & 0.439 & -0.109 \\
      \hline
    \end{tabular}%
  }
\end{table}

These difficult windows coincide with significant shifts in the vulnerability
landscape. The 2019\_M05--06 period includes multiple processor-related
vulnerabilities following Spectre/Meltdown, introducing new classes, while
2020\_M05--06 corresponds to early pandemic digitalisation, exposing new
attack surfaces. Selective replay methods (Hybrid-CASR, CASR) remain resilient
during these periods, whereas strong regularisation (OLoRA) and extended
replay (Replay-3P) can degrade performance, suggesting that over-constraining
the model hampers necessary adaptation.

\section{\uppercase{Discussion}}
\label{sec:discussion}

\subsection{Interpretation of Key Findings}
\label{sec:disc-interpretation}

\noindent
The experiments show that different temporal granularities achieve similar mean
Macro-F1 scores, from 0.651 (bi-monthly) to 0.669 (6- and 12-month windows).
This challenges implicit assumptions in temporal evaluation literature, where
data splitting is known to affect performance~\citep{Lyu2022Splitting} but certain
window sizes are often presumed to be clearly superior.

The relatively uniform performance can be interpreted through the lens of
concept drift. \citet{Hinder2024ConceptDrift} distinguish between
sudden and gradual drift patterns; both are present in the dataset, with gradual
changes within stable periods and sharp transitions, notably the 2021--2022
jump from roughly 1{,}000 to more than 2{,}000 disclosures per period. Different
granularities appear to capture these dynamics with comparable effectiveness.

Cumulative training yields only a modest improvement (F1~$=0.661$) over the
window-only baseline (F1~$=0.651$), despite a 15.9$\times$ increase in training
time. This contradicts the common expectation that more data systematically
leads to better models and aligns with \citet{Morrison2015HotSOS} observation that
additional historical data does not necessarily improve vulnerability
prediction. The backward retention results highlight a paradox: cumulative training offers
perfect stability (0\% decay) but lower absolute retention (IBR@1~$=0.661$) than replay-based methods such as
Replay-1P (0.791) and Hybrid-CASR (0.741), suggesting that exhaustive memory
preservation may impair effective adaptation.

\subsection{Theoretical Implications}
\label{sec:disc-theory}

\noindent
The performance of Hybrid-CASR emphasises the importance of selective memory
in code-based continual learning. Maximally interfered retrieval originally
proposed focusing rehearsal on uncertain examples~\citep{Aljundi2019MIR}, and
this principle transfers partially to vulnerability detection. However, the
domain exhibits highly variable class ratios (15--60\% positives across windows),
which makes pure uncertainty-based selection unstable.

Empirically, standard iCaRL-style exemplar replay~\citep{Rebuffi2017iCaRL}
proved brittle during high-drift periods such as 2020\_M05--06. Confidence-based
selection improves robustness, but without explicit class balancing the model can
lose minority-class patterns. Hybrid-CASR therefore combines uncertainty-driven
selection with a reserved portion of buffer capacity for class balancing.

The relative ranking among replay methods is also informative. Replay-1P
achieves the highest backward retention (IBR@1~$=0.791$) while maintaining
strong forward performance (F1~$=0.659$), whereas Replay-3P (F1~$=0.622$)
underperforms despite retaining more historical data. This contradicts the
assumption that larger buffers monotonically improve performance~\citep{Han2023CLCompare}
and suggests that, in rapidly evolving domains, controlled forgetting can be
more useful than comprehensive retention.

OLoRA's relatively poor performance (F1~$=0.599$) raises questions about the
applicability of strict orthogonality constraints in code vulnerability
detection. While such constraints have been successful in vision
tasks~\citep{LopezPaz2017GEM} and parameter-efficient continual learning for
LLMs~\citep{Song2023ConPET}, they appear overly rigid when vulnerability types overlap across time.

\subsection{Practical Implications}
\label{sec:disc-practical}

\noindent
The resource--performance analysis has direct implications for deployment.
Hybrid-CASR achieves F1~$=0.667$ in 432\,s per bi-monthly window, which makes
frequent retraining feasible on a single A100 GPU while mitigating model
staleness~\citep{Lyu2022Splitting}. Its memory overhead (19{,}847\,MB vs.\
13{,}525\,MB for window-only) remains manageable and is consistent with the
philosophy of parameter-efficient methods such as LoRA and
QLoRA~\citep{Hu2021LoRA,Dettmers2023QLoRA}. For environments with tighter
resource constraints, window-only training (F1~$=0.651$, 520\,s, 13{,}525\,MB)
provides a reasonable baseline. In contrast, cumulative training (138.2 minutes
per window for F1~$=0.661$) is difficult to justify given its cost.

Performance variability across windows (Hybrid-CASR F1 in the range 0.464--0.728)
is consistent with earlier observations of unstable vulnerability
prediction~\citep{Jimenez2016VPM}. The hardest windows
(2019\_M05--06, 2020\_M05--06), where the window-only baseline reaches only
0.539 and 0.548 F1 respectively, coincide with significant shifts in the
vulnerability landscape. All methods degrade in these periods, underscoring that human
expertise remains essential during major regime changes~\citep{Morrison2015HotSOS}.

\subsection{Threats to Validity}
\label{sec:disc-limitations}

\noindent
Several methodological choices limit the interpretation of the results. Macro-F1
treats classes symmetrically, which may misalign with operational settings where
false negatives carry higher cost~\citep{Davis2006PRROC,Saito2015PRBetter}.
Function-level granularity may miss vulnerabilities that span multiple functions
or files~\citep{Shin2011Complexity}. The CVE-to-commit linking through
CVEfixes~\citep{Bhandari2021CVEfixes} trades precision for scale compared to
manual curation~\citep{Ponta2019Dataset}, introducing label noise~\citep{Morrison2015HotSOS}.

The study focuses on a single architecture (phi-2), which constrains conclusions
about model-size and architecture effects. \citet{Konno2025Taltech}
demonstrated that decoder LLMs can be effective across languages and code
granularities, but continual learning behaviour may differ for encoder models
(CodeBERT~\citep{Feng2020CodeBERT}) or graph-based approaches~\citep{Zhou2019Devign,Cao2021BGNN4VD}. The 2018--2024
time span captures recent trends but cannot guarantee future applicability.

Using phi-2 exclusively also raises a validity threat related to
pretraining data. Phi-2 was released in December 2023, and its pretraining
corpus likely includes public repositories containing vulnerabilities from
the 2018--2023 evaluation period. This temporal overlap implies that the model
may have seen similar vulnerable code during pretraining,
potentially inflating performance. A more rigorous setup would either employ models whose
pretraining cutoff predates the evaluation window or restrict evaluation
to vulnerabilities disclosed after the model's training date.

The CVEfixes-based dataset predominantly covers C/C++ and Java, which may bias
results towards these languages. The fixed
decision threshold (0.5) assumes symmetric error costs; in practice,
organisations may favour high recall or high precision depending on review
capacity and risk tolerance.

\subsection{Future Research Directions}
\label{sec:disc-future}

\noindent
The near-constant performance across temporal granularities (F1 0.651--0.669)
should be validated on other datasets to determine whether it is a general
phenomenon. One direction is adaptive segmentation, where window sizes are
adjusted based on drift-detection signals~\citep{Hinder2024ConceptDrift}. The
success of selective replay motivates exploring alternative selection
strategies, such as gradient-based importance measures~\citep{Aljundi2019MIR} or
diversity-aware sampling, and more flexible knowledge-organisation mechanisms,
including adapter fusion~\citep{Pfeiffer2021AdapterFusion}.
Finally, future work should develop evaluation
protocols for zero-day-like settings through cross-family evaluations~\citep{Lu2021CodeXGLUE,Morrison2015HotSOS}.

\section{\uppercase{Conclusion}}
\label{sec:conclusion}

\noindent
This paper examined whether continual learning strategies can mitigate the
temporal challenges of vulnerability detection when large language models are
deployed over multiple years of evolving code. Using a CVEfixes-derived dataset
spanning 42 bi-monthly windows from 2018 to 2024 and a LoRA-tuned
\texttt{phi-2} model, we evaluated temporal granularity, forgetting mitigation
mechanisms and computational trade-offs under a strict forward-chaining
protocol.

The proposed Hybrid-CASR method, which combines confidence-based example
selection with explicit class balancing, achieves the best overall forward
performance with a mean Macro-F1 of 0.667, a 0.016 absolute gain over the
window-only baseline (0.651). The improvement is statistically significant
(Wilcoxon signed-rank test, $p = 0.026$) but small in magnitude (Cliff's
$\delta = 0.103$), indicating incremental rather than transformative benefits.
Hybrid-CASR attains this performance with only modest computation compared to
simple per-window fine-tuning and is far cheaper than fully cumulative training:
its F1-per-minute efficiency is around 24\% higher than the window-only baseline,
while cumulative training over all past windows provides only minimal gains
(Macro-F1~$=0.661$) at a 15.9-fold increase in training time and backward
retention (IBR@1~$=0.661$) that remains below selective replay methods such as
Replay-1P (0.791) and Hybrid-CASR (0.741).

Temporal granularity analysis shows that models are surprisingly robust to
window size: granularities from one to twelve months yield similar mean F1
scores between 0.651 and 0.669. The choice of granularity appears to affect
\emph{which} vulnerabilities are captured more than aggregate detection
capability.

From an operational perspective, these findings imply that current learning-based
vulnerability detectors are best deployed as decision-support tools rather than
stand-alone oracles. Forward performance in the 65--67\% Macro-F1
range means human verification remains essential. The limited performance gap
between strategies indicates that organisations with limited ML capacity may reasonably
adopt simple per-window fine-tuning, while teams with
additional resources can use Hybrid-CASR for a slightly better
accuracy--efficiency trade-off.

This study has several limitations: it focuses on a single decoder-only
architecture; the dataset primarily covers C/C++ and Java; and potential pretraining contamination may blur memorisation and temporal
generalisation. Future work should validate findings across different architectures and datasets, investigate adaptive windowing strategies, and
design evaluation protocols that better approximate zero-day scenarios.

Overall, this work provides a systematic temporal evaluation of LLM-based
vulnerability detection and quantifies what contemporary continual learning
strategies can and cannot deliver. Hybrid-CASR offers a small
but measurable improvement over simpler baselines, but the modest effect sizes and persistent
performance variability underline that
robust temporal vulnerability detection remains an open research challenge.



\bibliographystyle{apalike}
{\small
\bibliography{example}}

\end{document}